\documentclass[11pt]{article} 
\usepackage[toc,page]{appendix}
\usepackage[utf8]{inputenc} 

\usepackage{geometry} 
\usepackage{graphicx} 

\usepackage{amsmath}
\usepackage{bm} 
\usepackage{bbm}
\usepackage[english]{babel}
\usepackage{natbib} 
\usepackage{amsfonts}
\usepackage[strict]{changepage}
\usepackage{hyphenat}
\usepackage{booktabs} 
\usepackage{enumerate}
\usepackage{kbordermatrix} 
\usepackage{array} 
\usepackage{paralist} 
\usepackage[group-separator={,}]{siunitx}
\usepackage{verbatim} 
\usepackage{subfig} 
\usepackage{setspace} 
\usepackage{caption}
\usepackage{lscape}
\usepackage{tabu}
\usepackage[flushleft]{threeparttable}

\usepackage{fancyhdr} 
\usepackage{sectsty}

\usepackage[nottoc,notlof,notlot]{tocbibind} 
\usepackage[titles,subfigure]{tocloft} 

\title{\textbf{The Generalised Isolation-With-Migration Model: a Maximum-Likelihood Implementation for Multilocus Data Sets}}
\author{Rui J. Costa and Hilde Wilkinson-Herbots}
\author{\small Rui J. Costa and Hilde Wilkinson-Herbots\\
\small Department of Statistical Science, \small University College London\\
\small Gower Street, London WC1E 6BT, UK\\
\small email: rui.costa.11@ucl.ac.uk
}
\date{} 

\begin{document}
\maketitle

\begin{abstract}
Statistical inference about the speciation process has often been based on the isolation-with-migration (IM) model, especially when the research aim is to learn about the presence or absence of gene flow during divergence. The generalised IM model introduced in this paper extends both the standard two-population IM model and the isolation-with-initial-migration (IIM) model, and encompasses both these models as special cases. It can be described as a two-population IM model in which migration rates and population sizes are allowed to change at some point in the past. By developing a maximum-likelihood implementation of this GIM model, we enable inference on both historical and contemporary rates of gene flow between two closely related species. Our method relies on the spectral decomposition of the coalescent generator matrix and is applicable to data sets consisting of the numbers of nucleotide  differences between one pair of DNA sequences at each of a large number of independent loci. 
\end{abstract}

\begin{adjustwidth}{3.5em}{3.5em}
\end{adjustwidth}

\section{Introduction}
Coalescent-type stochastic models can be used as a statistical inference tool to extract information from a sample of genomic sequences. When the aim is to learn about the role of gene flow during speciation,  most inferential methods are based on the isolation-with-migration (IM) model  \citep[see, e.g.,][]{Nielsen2001,Hey2004,Hey2005,Hey2007, Hey2010}. A survey of research that has used the IM model in the context of speciation can be found in \citet{Pinho2010}. In recent years, as more extensions of the IM model became available, some authors have taken on the task of finding the evolutionary scenario, represented by some version of the IM model, that best explains a given polymorphism data \citep[see, e.g.,][]{Wang2010,Lohse2011,Lohse2014}. 

\begin{figure}[h] 
\graphicspath{ {./} }
\centering         
\includegraphics[width=7.5cm, angle=0]{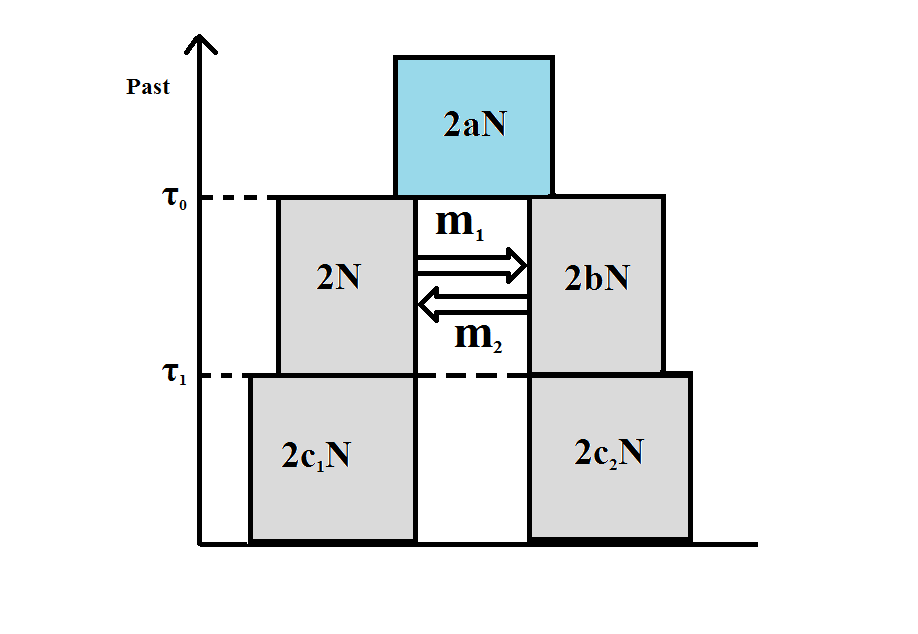} 
\vspace*{-0.25cm}     
\caption{The isolation-with-initial-migration (IIM)  model \citep{Herbots2012,Costa2016}. Population size parameters $a$, $b$, $c_{1}$, and $c_{2}$ are in units of $2N$ sequences, where $N$ is the effective population size of the species on the left of the diagram, during the migration stage. From a forward-in-time perspective, $\tau_{0}$ denotes the splitting time of the ancestral population and the beginning of the gene flow stage; after $\tau_{1}$, gene flow ceases. The rates of gene flow are represented by $m_{1}$ and $m_{2}$.}      
\label{fig:IIMfull1} 
\end{figure}

\begin{figure}[h] 
\graphicspath{ {./} }
\centering         
\includegraphics[width=10.5cm, angle=0]{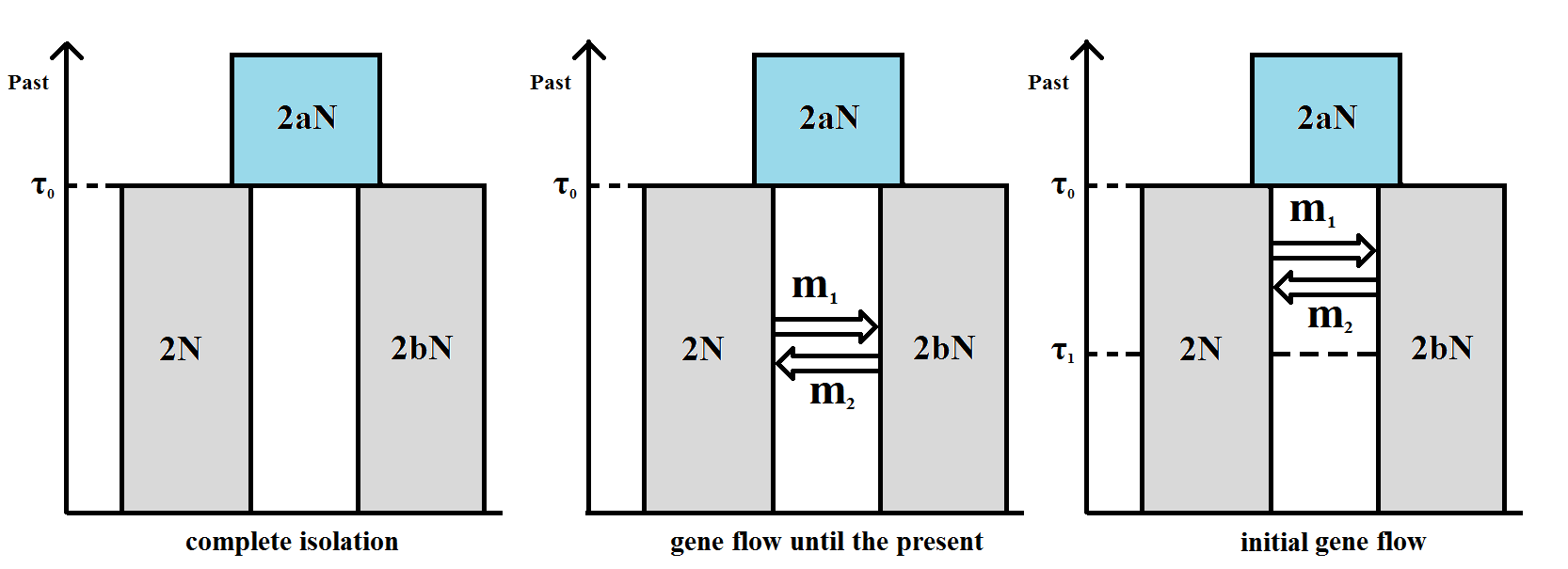} 
\vspace*{-0.25cm}     
\caption{Three models of divergence nested in the isolation-with-initial-migration (IIM) model. The parameters have the same meaning as in Figure \ref{fig:IIMfull1}.}      
\label{fig:IIMnested} 
\end{figure}

A recent addition to the list of implementable IM models is the so-called isolation-with-initial-migration (IIM) model \citep{Herbots2012,Herbots2015,Costa2016}. This is a 2-population IM model in which gene flow may stop at some point in the past (see Figure \ref{fig:IIMfull1}). As a result of this development, it is now possible to assess which of three divergence scenarios is most supported by a given data set: divergence without gene flow, divergence with constant gene flow until the present, or divergence with initial gene flow and subsequent isolation. In fact, one way to perform this comparison is to fit the three models depicted in Figure \ref{fig:IIMnested}: a complete isolation model, a standard IM model, and a version of the IIM model in which the sizes of the diverging populations are kept constant. The aim of this latter restriction is to separate, as much as possible, the effect of allowing for different gene flow scenarios from the effect of allowing for population size changes.

\begin{figure}[h] 
\graphicspath{ {./} }
\centering         
\includegraphics[width=10.5cm, angle=0]{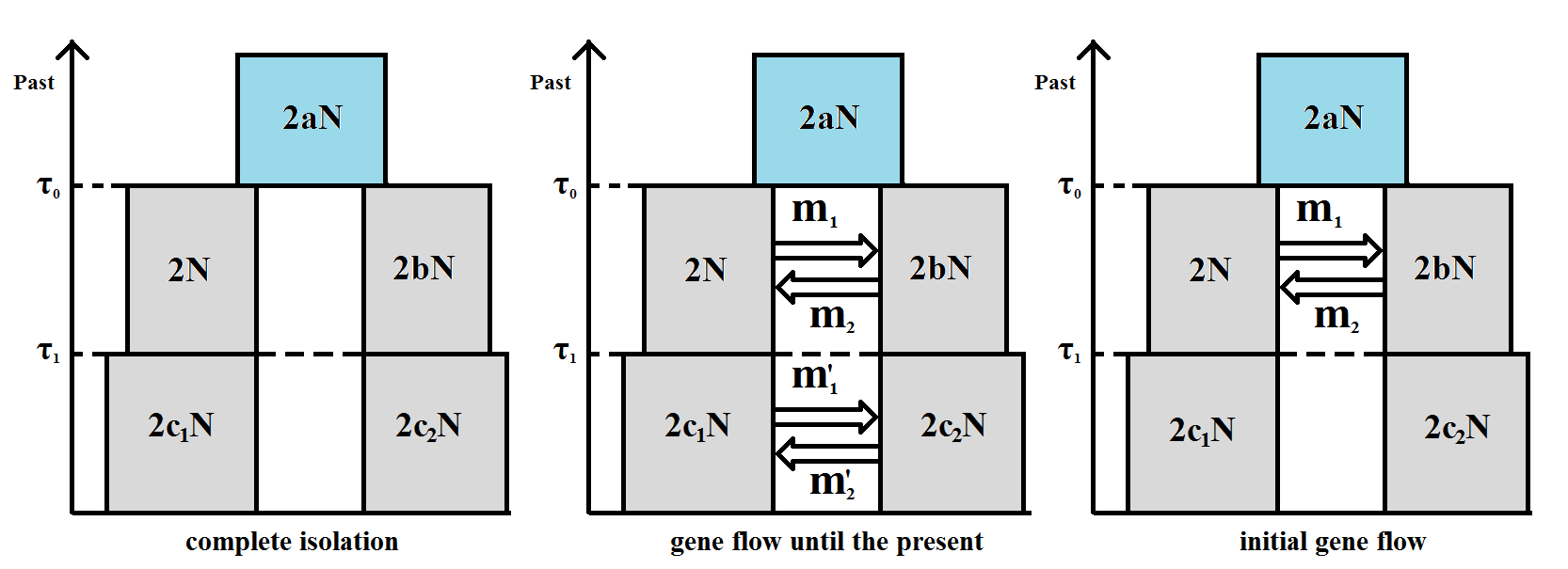} 
\vspace*{-0.25cm}     
\caption{The full GIM model (centre) and two models of divergence nested in it. The parameters $m'_{1}$ and $m'_{2}$ in the full GIM model denote the rates of contemporary gene flow.  The remaining parameters have the same meaning as in Figure \ref{fig:IIMfull1}.}      
\label{fig:GIM} 
\end{figure}

In practice, however, one is often ignorant of whether the sizes of the populations during divergence have changed significantly or not, and allowing for population size changes may improve the fit of the models substantially. Therefore, we would like to be able to compare  the three gene flow scenarios in a framework which incorporates the full IIM model shown in Figure \ref{fig:IIMfull1}. The aim of this paper is to build such a framework, by developing a maximum-likelihood implementation of a model which we call the \textit{generalised isolation-with-migration} (GIM) model. This will enable us to compare the three models shown in Figure \ref{fig:GIM}, which include the full GIM model (central diagram) and two models nested in it. More specifically, our goal is to enable these models to be fitted to data sets consisting of observations on the number of segregating sites between pairs of DNA sequences from a large number of independent, non-recombining loci.  

This paper follows a series of papers on estimation methods  which are based on explicit likelihood expressions and are suited for multilocus data sets. The likelihood of the number of pairwise differences under the IM model was derived  in \citet{Herbots2008} and later extended to the isolation-with-initial-migration (IIM) model in \citet{Herbots2012} and \citet{Costa2016}. The results of  \citet{Lohse2011} for the IM model included the likelihood of data on triplets and are based on the solution of systems of generating functions. Making use of spectral decomposition and lumpability of continuous-time Markov chains, \citet{Andersen2014} obtained explicit results for an IM model with an arbitrary number of lineages in
an arbitrary number of populations. \citet{Lohse2014} derived the likelihood of full mutational configurations of sequences under both
admixture and ancestral structure scenarios.

\section{Theory and methods}

From a backward-in-time perspective, the fullest GIM model we consider consists of two successive 2-island models and one ancestral Wright-Fisher population, as illustrated in the central diagram of Figure \ref{fig:GIM}. The population on the left of the diagram will be referred to as `population 1' and the population on the right as `population 2'. The time parameters $\tau_{1}>0$ and $\tau_{0}>\tau_{1}$ are in units of $2N$ generations, where $2N$ is the number of haploid genomes in population 1 during the second stage of migration. The relative sizes of the remaining populations with respect to the size of population 1 between $\tau_{1}$ and $\tau_{0}$ are given by the parameters $a$, $b$, $c_{1}$ and $c_{2}$. The parameters $m_{i}$ and $m'_{i}$, with $i \in \left\lbrace 1,2 \right\rbrace$, represent the backward migration rates from population $i$ to $j$ ($i \neq j$) per generation, i.e. the fraction of population $i$ which migrates to population $j$ in each generation. The reproduction in each population follows the neutral Wright-Fisher model. It is assumed that, in each generation, the process of reproduction restores the population to its original size, in case the number of immigrants is different from the number of emigrants.  All parameters of the GIM model are strictly positive, except for the migration rates, which are non-negative.

We are interested in the genealogical process of a random sample of two DNA sequences at the same locus, taken from either of the present populations (or one from each population), under the GIM model. This process is a succession of discrete-time Markov chains tracing the lineages ancestral to the sample back in time. It is absorbed whenever the two lineages coalesce at their most recent common ancestor. The process can start in one of three states: if both sequences are sampled from population 1, the initial state is `1'; if both come from population 2, or there is one from each population,  the initial states are denoted `2' and `3' respectively. Until time $\tau_{0}$ into the past, the process is either in one of these three states or coalescence has occurred (state `4'). After $\tau_{0}$, only two situations are possible: either there are two distinct ancestral lineages (states `1', `2' and `3'), or coalescence has occurred (state `4').

The genealogy of the sample under the GIM model is a stochastic process that runs in discrete time. But if time is measured in units of $2N$ generations and $N$ is large, it is well approximated by \textit{the coalescent under the GIM model}, which is composed of three consecutive continuous-time Markov chains \citep{Kingman1982coal,Notohara1990}.

\subsection{The coalescent under the GIM model}

The coalescent under the GIM model is defined by the following generator matrices. For $0 \leq t \leq \tau_{1}$,
\small
\begin{equation}
\label{matrix:Q1}
\renewcommand{\arraystretch}{1.5}
\mathbf{Q_{1}}=\kbordermatrix{~&(1)&(3)&(2)&(4)\\
				(1)&-\left(\frac{1}{c_{1}}+M'_{1}\right)&M'_{1}&0&\frac{1}{c_{1}}\\
				(3)&\frac{M'_{2}}{2}&-\left(\frac{M'_{1}+M'_{2}}{2}\right)&\frac{M'_{1}}{2}&0\\
			    (2)&0&M'_{2}&-\left(\frac{1}{c_{2}}+M'_{2}\right)&\frac{1}{c_{2}}\\
			    (4)&0&0&0&0}\, 
\end{equation} \normalsize
\citep{Notohara1990}, where \small $M'_{i}/2=2N m'_{i}$ \normalsize is the rate of migration of a single lineage when in population $i$ ($i \in \{1,2\}$).  The rate \small $\frac{1}{c_{i}}$ \normalsize is the rate of coalescence of two lineages if both are in population $i$. Note that, for mathematical convenience, state 2 corresponds to row and column three, whereas state 3 corresponds to row and column 2: this makes $\mathbf{Q_{1}}$ as symmetric as possible, while reserving states 1 and 2 for the states in which two lineages are present in population 1 and population 2 respectively.  If $\tau_{1}< t \leq \tau_{0}$, \small
\begin{equation}
\label{matrix:Q2}
\renewcommand{\arraystretch}{1.5}
\mathbf{Q_{2}}=\kbordermatrix{~&(1)&(3)&(2)&(4)\\
				(1)&-\left(1+M_{1}\right)&M_{1}&0&1\\
				(3)&\frac{M_{2}}{2}&-\left(\frac{M_{1}+M_{2}}{2}\right)&\frac{M_{1}}{2}&0\\
			    (2)&0&M_{2}&-\left(\frac{1}{b}+M_{2}\right)&\frac{1}{b}\\
			    (4)&0&0&0&0}\, ,
\end{equation} \normalsize
where $1$ and \small $\frac{1}{b}$ \normalsize are the coalescence rates of two lineages in population 1 and population 2 respectively, and \small $M_{i}/2= 2N m_{i}$\normalsize. Finally, for $t > \tau_{0}$, 
\small
\begin{equation}
\label{matrix:Q3}
\renewcommand{\arraystretch}{1.5}
\mathbf{Q_{3}}=\kbordermatrix{~&(1)&(3)&(2)&(4)\\
				(1)&-\frac{1}{a}&0&0&\frac{1}{a}\\
				(3)&0&-\frac{1}{a}&0&\frac{1}{a}\\
				(2)&0&0&-\frac{1}{a}&\frac{1}{a}\\ 
				(4)&0&0&0&0}\, 
\end{equation}\normalsize
\citep{Kingman1982coal}, where $\frac{1}{a}$ is the rate of coalescence of two lineages in the ancestral population.

The matrix of transition probabilities $\mathbf{P}(t)$ of the coalescent under the GIM model has the following form:
\small
\begin{equation}
\label{eq:tran_prob_matrix}
\begin{array}{lcl}

\mathbf{P}(t) &=& \left\{
  \begin{array}{l l}
    e^{\mathbf{Q_{1}}t}\quad& \quad \text{for } 0 \leq t \leq \tau_{1},\\
   e^{\mathbf{Q_{1}}\tau_{1}}\, e^{\mathbf{Q_{2}}\left(t-\tau_{1}\right)} \quad  & \quad \text{for } \tau_{1} < t \leq \tau_{0},\\
   e^{\mathbf{Q_{1}}\tau_{1}}\, e^{\mathbf{Q_{2}}\left(\tau_{0}-\tau_{1}\right)}\,e^{\mathbf{Q_{3}}\left(t-\tau_{0}\right)}\quad & \quad \text{for }\tau_{0} < t< \infty,\\
0\quad &\quad \text{otherwise}.
  \end{array} \right.\\
  \end{array}
\end{equation}\normalsize
Recall that all time and population size parameters are assumed strictly positive. In Section 2.1 of \citet{Costa2016}, we prove that, if both migration rates are also strictly positive, the matrices $\mathbf{Q_{1}}$ and $\mathbf{Q_{2}}$ are diagonalisable and have non-positive real eigenvalues. Moreover, the matrix
\small
\begin{equation}
\label{eq:eigenmat}
\begin{array}{c}
\begin {bmatrix} 
1 &0&1&0\\ 
1 &1&0&0\\
1& 0&0&1\\
1&0&0&0
\end{bmatrix}
\end{array}
\end{equation}
\normalsize
contains a set of four independent right eigenvectors of $\mathbf{Q_{3}}$, and  the corresponding vector of eigenvalues is \small $\left(0,-1/a, -1/a, -1/a\right)\,$. \normalsize 
 Hence, for \small $M_{1}, M_{2},M'_{1}, M'_{2} >0$, $\mathbf{P}(t)$ \normalsize can be written as:
\small
\begin{equation}
\label{eq:p(t)_decomposition}
\begin{array}{lcl}

 \mathbf{P}(t) &=&\left\{
  \begin{array}{l l}
    \mathbf{G^{-1}}e^{\mathbf{-A}t}\mathbf{G}\quad& \quad \text{for } 0 \leq t \leq \tau_{1},\\
   \mathbf{G^{-1}}e^{\mathbf{-A}\tau_{1}}\mathbf{G}\,\mathbf{C^{-1}}e^{\mathbf{-B}(t-\tau_{1})}\mathbf{C}  \quad  & \quad \text{for } \tau_{1} < t \leq \tau_{0},\\
   \mathbf{G^{-1}}e^{\mathbf{-A}\tau_{1}}\mathbf{G}\,\mathbf{C^{-1}}e^{\mathbf{-B}(\tau_{0}-\tau_{1})}\mathbf{C}\, \mathbf{D^{-1}}e^{\mathbf{-\Gamma}(t-\tau_{0})}\mathbf{D} \,  & \quad \text{for } \tau_{0} < t< \infty,\\
0\quad &\quad \text{otherwise},
  \end{array} \right.
  \end{array}
\end{equation} \normalsize
where $\mathbf{G}$, $\mathbf{C}$ and $\mathbf{D}$ are the matrices of right eigenvectors of $\mathbf{Q_{1}}$, $\mathbf{Q_{2}}$ and $\mathbf{Q_{3}}$ respectively, and $\mathbf{-A}$, $\mathbf{-B}$ and $\boldsymbol{-\Gamma}$ are the corresponding diagonal matrices of (non-positive, real) eigenvalues. The entries in the main diagonals of $\mathbf{A}$, $\mathbf{B}$ and $\mathbf{\Gamma}$ contain the absolute values of the eigenvalues, and  are represented by the letters $\alpha_{i}=\mathbf{(A)}_{ii}$, $\beta_{i}=\mathbf{(B)}_{ii}$ and $\gamma_{i}=\mathbf{(\Gamma)}_{ii}$.

If a matrix $\mathbf{Q}$ is a generator matrix of a migration stage in the GIM model, with migration parameters \small $M_{i}=M_{j}=0$  ($i, j \in \left\lbrace 1,2 \right\rbrace$ \normalsize and \small $i \neq j$) \normalsize and relative population size parameters $c_{i}$ and $c_{j}$, then its right eigenvectors are those shown in matrix (\ref{eq:eigenmat}) and  its vector of eigenvalues is \small $\left( 0, 0, -1/c_{1}, -1/c_{2} \right)\,$. \normalsize So when there is no gene flow between $\tau_{0}$ and $\tau_{1}$, or no gene flow between $\tau_{1}$ and the present, $\mathbf{P}(t)$ can still be decomposed as in equation (\ref{eq:p(t)_decomposition}). 

For all values of $M_{1}$ and $M_{2}$, the characteristic polynomial of $\mathbf{Q}$, denoted $\mathcal{P}_{\mathbf{Q}}(\beta)$, is of the form $\beta \times \mathcal{P}_{\mathbf{Q^{\left(r\right)}}}(\beta)$, where $\mathbf{Q^{\left(r\right)}}$ is the three by three upper-left submatrix of $\mathbf{Q}$. So $\mathbf{Q}$ has a zero eigenvalue and its remaining eigenvalues are the eigenvalues of $\mathbf{Q}^{(r)}$. If $\mathbf{Q}$ has migration parameters \small $M_{i}=0$ \normalsize and \small $M_{j}>0$, \normalsize $\mathbf{Q}^{(r)}$ becomes triangular. The eigenvalues of $\mathbf{Q}^{(r)}$ will be the entries in its main diagonal. Hence the vector of eigenvalues of $\mathbf{Q}$ will be $\boldsymbol{\lambda}=\left[-1/c_{i}, -M_{j}/2, -(M_{j}+1/c_{j} ), 0 \right]$. If there are no repeated eigenvalues in $\boldsymbol{\lambda}$, we can be sure that $\mathbf{Q}$ is diagonalisable (and its eigenvalues are non-positive and real). In other words, even if there is unidirectional migration between $\tau_{1}$ and the present, or between $\tau_{0}$ and $\tau_{1}$, the probability transition matrix \small $\mathbf{P}(t)$ \normalsize can still be decomposed as in (\ref{eq:p(t)_decomposition}), as long as there are no repeated entries in $\boldsymbol{\lambda}$. Two comments are in order here: first, repeated eigenvalues will occur if and only if \small $1/c_{i} = M_{j}/2 $ \normalsize or \small $1/c_{i} = M_{j}+1/c_{j}$\normalsize; second, the set of parameter values that make these equalities true is negligible when compared to the whole parameter space, so it is very unlikely that the likelihood maximisation procedure chooses values from this set (although one should be careful to avoid using them as initial values).

The probability that, starting in state $i$ ($i \in \{1,2,3\}$), the process has reached state 4 by time $t$ is given by the entry corresponding to the $i^{\mathrm{th}}$ row and 4$^{\mathrm{th}}$ column of $\mathbf{P}(t)$. This is also the cumulative distribution function (\textit{cdf}) of $T_{i}$, the time until coalescence, which we denote $F_{T_{i}}(t)$. If the initial state is $i$, and $p^{(1)}_{ij}(t)$, $p^{(2)}_{jl}(t)$ and $p^{(3)}_{l4}(t)$ denote transition probability functions of the Markov chains with generator matrices $\mathbf{Q_{1}}$, $\mathbf{Q_{2}}$ and $\mathbf{Q_{3}}$ respectively, then:
\small
\begin{equation}
\label{eq:cdf_T_1}
F_{T_{i}}(t) = \left\{
  \begin{array}{l l}
   p^{(1)}_{i4}(t) & \quad \text{for }0 \leq t \leq \tau_{1},\\
   \\
    \displaystyle\sum_{j=1}^{4}p^{(1)}_{ij}(\tau_{1})\,p^{(2)}_{j4}(t-\tau_{1}) \quad  & \quad \text{for } \tau_{1} < t \leq \tau_{0},\\
    \\
 \displaystyle\sum_{j=1}^{4} p^{(1)}_{ij}(\tau_{1})\displaystyle\sum_{l=1}^{4}p^{(2)}_{jl}(\tau_{0}-\tau_{1})\,p^{(3)}_{l4}(t-\tau_{0}) \quad& \quad\text{for } \tau_{0} < t< \infty,\\
   \\
0\quad &\quad \text{otherwise}.
  \end{array} \right.
\end{equation}
\normalsize
Representing by $A_{mn}$ the $(m,n)$ entry of a matrix $\mathbf{A}$, and by $A^{-1}_{mn}$ the same entry of the matrix $\mathbf{A^{-1}}$, we have that \small $p^{(1)}_{ij}(t)=\sum_{k=1}^{4} G_{ik}^{-1}G_{kj}\,e^{-\alpha_{k}t}$, $p^{(2)}_{ij}(t)=\sum_{k=1}^{4} C_{ik}^{-1}C_{kj}\,e^{-\beta_{k}t}$  \normalsize and \small $p^{(3)}_{i4}(t)=\sum_{k=1}^{4} D_{ik}^{-1}D_{k4}\,e^{-\gamma_{k}t}$ \normalsize.

Differentiating the expression above gives the following density for $T_{i}\,$:
\small
\begin{equation}
\label{eq:pdf_T_1}
f_{T_{i}}(t) = \left\{
  \begin{array}{l l}
   f^{(1)}_{i}(t) & \quad \text{for }0 \leq t \leq \tau_{1},\\
   \\
    \displaystyle\sum_{j=1}^{4}p^{(1)}_{ij}(\tau_{1})\,f^{(2)}_{j}(t-\tau_{1}) \quad  & \quad \text{for }\tau_{1} < t \leq \tau_{0},\\
    \\
 \displaystyle\sum_{j=1}^{4} p^{(1)}_{ij}(\tau_{1})\displaystyle\sum_{l=1}^{4}p^{(2)}_{jl}(\tau_{0}-\tau_{1})\,f^{(3)}_{l}(t-\tau_{0}) \quad& \quad \text{for } \tau_{0} < t< \infty,\\
   \\
0\quad, &\quad \text{otherwise},
  \end{array} \right.
\end{equation}
\normalsize
where \small $f^{(1)}_{i}(t)=\sum_{k=1}^{4} -\alpha_{k}\,G_{ik}^{-1}G_{k4}\,e^{-\alpha_{k}t}$, $f^{(2)}_{i}(t)=\sum_{k=1}^{4}-\beta_{k}\, C_{ik}^{-1}C_{k4}\,e^{-\beta_{k}t}$ \normalsize and  \small $f^{(3)}_{i}(t)=\sum_{k=1}^{4}-\gamma_{k}\, D_{ik}^{-1}D_{k4}\,e^{-\gamma_{k}t}$. \normalsize

\subsection{The distribution of the number of pairwise nucleotide differences}
We assume the infinite sites model of \citet{Watterson1975}, according to which: a) in each generation, the number of mutations occurring in a sequence at a particular locus follows a Poisson distribution with mean $\mu$; and b) no two mutations ever occur at the same nucleotide site. In the coalescent approximation (measuring time in units of $2N$ generations), mutations accumulate on a pair of lineages according to a Poisson process of rate $\theta=4N\mu$ ($\theta$ is the scaled mutation rate at the locus considered). 
Given the coalescence time $T_{i}$ of two DNA sequences at this locus, their number of segregating sites $S_{i}$ is Poisson distributed with mean $\theta T_{i}$.
 Denoting $g_{s}(t):=\frac{(\theta t)^{s}}{s!}e^{-\theta t}$, we have, for $s \in \{0,1,2,...\}$,

\small
\begin{equation}
\begin{array}{lcl}
\mathrm{P}(S_{i}=s)&=&\mathrm{E}[g_{s}(T_{i})]\\
\\
&=&\displaystyle\int_{0}^{\tau_{1}} \! g_{s}(t) \, f^{(1)}_{i}(t) \mathrm{d}t + \displaystyle \sum_{j=1}^{4}\,p^{(1)}_{ij}(\tau_{1})\,\int_{\tau_{1}}^{\tau_{0}} \!g_{s}(t)\,f_{j}^{(2)}\left(t-\tau_{1}\right)\mathrm{d}t\\
\\
&&+  \displaystyle\sum_{j=1}^{4} p^{(1)}_{ij}(\tau_{1})\displaystyle\sum_{l=1}^{4}p^{(2)}_{jl}(\tau_{0}-\tau_{1})\,\int_{\tau_{0}}^{\infty} \!g_{s}(t)\,f^{(3)}_{l}(t-\tau_{0})  \mathrm{d}t \quad,

\end{array}
\label{eq:likelihood_single_0}
\end{equation}
\normalsize
\noindent
where $i$ is again the initial state of the coalescent, corresponding to the sampling locations of the two sequences.
Changing the limits of integration, equation (\ref{eq:likelihood_single_0}) becomes: 
\small
\begin{equation*}
\begin{array}{lcl}
\mathrm{P}\left(S_{i}=s\right)&=&\displaystyle\int_{0}^{\tau_{1}} \! g_{s}(t) \, f^{(1)}_{i}(t) \mathrm{d}t + \displaystyle \sum_{j=1}^{4}\,p^{(1)}_{ij}(\tau_{1})\,\int_{0}^{\tau_{0}-\tau_{1}} \!g_{s}(\tau_{1}+t)\,f_{j}^{(2)}\left(t\right)\mathrm{d}t\\
\\
&& +\displaystyle\sum_{j=1}^{4} p^{(1)}_{ij}(\tau_{1})\displaystyle\sum_{l=1}^{4}p^{(2)}_{jl}(\tau_{0}-\tau_{1})\,\int_{0}^{\infty} \!g_{s}(\tau_{0}+t)\,f^{(3)}_{l}(t)  \mathrm{d}t \quad.\\

\end{array}
\end{equation*}
\normalsize
Denoting by $W_{i}$, $Y_{j}$ and $Z_{l}$ the random variables with \textit{pdf}'s $f_{i}^{(1)}$, $f_{j}^{(2)}$ and $f_{l}^{(3)}$ respectively, the above equation can be written as:
\small
\begin{equation*}
\begin{array}{lcl}
\mathrm{P}\left(S_{i}=s\right)&=&\mathrm{E}[g_{s}(W_{i})|W_{i}\leq \tau_{1}]\mathrm{P}[W_{i}\leq \tau_{1}] \\
\\
&&+ \displaystyle \sum_{j=1}^{4}\,p^{(1)}_{ij}(\tau_{1})\,\mathrm{E}[g_{s}(\tau_{1}+Y_{j})|\tau_{1}+Y_{j}\leq \tau_{0}]\,\mathrm{P}[\tau_{1}+Y_{j}\leq \tau_{0}]\\
\\
&&+  \displaystyle\sum_{j=1}^{4} p^{(1)}_{ij}(\tau_{1})\displaystyle\sum_{l=1}^{4}p^{(2)}_{jl}(\tau_{0}-\tau_{1})\,\mathrm{E}[g_{s}(\tau_{0}+Z_{l})] \quad.\\
\\
&=&\mathrm{E}[g_{s}(W_{i})]-\mathrm{E}[g_{s}(W_{i})|W_{i}> \tau_{1}]\mathrm{P}[W_{i}> \tau_{1}] \\
\\
&&+\displaystyle \sum_{j=1}^{4}\,p^{(1)}_{ij}(\tau_{1})\,\left\lbrace\mathrm{E}[g_{s}(\tau_{1}+Y_{j})]-\mathrm{E}[g_{s}(\tau_{1}+Y_{j})|\tau_{1}+Y_{j}> \tau_{0}]\,\mathrm{P}[\tau_{1}+Y_{j}>\tau_{0}]\right\rbrace\\
\\
&&+  \displaystyle\sum_{j=1}^{4} p^{(1)}_{ij}(\tau_{1})\displaystyle\sum_{l=1}^{4}p^{(2)}_{jl}(\tau_{0}-\tau_{1})\,\mathrm{E}[g_{s}(\tau_{0}+Z_{l})] \quad\\
\\

\end{array}
\end{equation*}
\normalsize
Recall that \small $f^{(1)}_{i}(t)=\sum_{k=1}^{4} -\alpha_{k}\,G_{ik}^{-1}G_{k4}\,e^{-\alpha_{k}t}$, $f^{(2)}_{i}(t)=\sum_{k=1}^{4}-\beta_{k}\, C_{ik}^{-1}C_{k4}\,e^{-\beta_{k}t}$ \normalsize and \linebreak \small $f^{(3)}_{i}(t)=\sum_{k=1}^{4}-\gamma_{k}\, D_{ik}^{-1}D_{k4}\,e^{-\gamma_{k}t}$,  \normalsize and that some eigenvalues of \small $\mathbf{Q_{1}}$, $\mathbf{Q_{2}}$ \normalsize and \small $\mathbf{Q_{3}}$ \normalsize are equal to zero, i.e. some of the $-\alpha_{k}$, $-\beta_{k}$ and $-\gamma_{k}$ are zero.  For those $\alpha_{k}$, $\beta_{k}$ and $\gamma_{k}$ that are strictly positive, we let $W^{*}_{k}$, $Y^{*}_{k}$ and $Z^{*}_{k}$ denote exponentially distributed random variables with rates $\alpha_{k}$, $\beta_{k}$ and $\gamma_{k}$ respectively.  The equation above can then be written as:
\small
\begin{equation*}
\begin{array}{lcl}
\mathrm{P}\left(S_{i}=s\right)&=&-\displaystyle\sum_{k:\alpha_{k}>0}\,G^{-1}_{ik}G_{k4}\left\lbrace\mathrm{E}[g_{s}(W^{*}_{k})]-\mathrm{E}[g_{s}(W^{*}_{k})|W^{*}_{k}> \tau_{1}]\mathrm{P}[W^{*}_{k}> \tau_{1}]\right\rbrace \\
\\
&&- \displaystyle \sum_{j=1}^{4}\,p^{(1)}_{ij}(\tau_{1})\, \displaystyle\sum_{k:\beta_{k}>0} C_{jk}^{-1}C_{k4}\left\lbrace\mathrm{E}[g_{s}(\tau_{1}+Y^{*}_{k})]\right.\\
\\
 &&\left.-\mathrm{E}[g_{s}(\tau_{1}+Y^{*}_{k})|\tau_{1}+Y^{*}_{k}> \tau_{0}]\,\mathrm{P}[\tau_{1}+Y^{*}_{k}>\tau_{0}]\right\rbrace\\
\\
&&-\displaystyle\sum_{j=1}^{4} p^{(1)}_{ij}(\tau_{1})\displaystyle\sum_{l=1}^{4}p^{(2)}_{jl}(\tau_{0}-\tau_{1})\,\displaystyle\sum_{k:\gamma_{k}>0} D_{lk}^{-1}D_{k4}\,\mathrm{E}[g_{s}(\tau_{0}+Z^{*}_{k})] \quad.\\
\\
\end{array}
\end{equation*}
\normalsize
Finally, making use of the lack of memory property of the exponential distribution gives:
\small
  \begin{equation}
  \label{eq:likelihood_single}
\begin{array}{lcl}
\mathrm{P}(S_{i}=s)&=&-\displaystyle\sum_{k:\alpha_{k}>0}G^{-1}_{ik}G_{k4}\left\lbrace\mathrm{E}[g_{s}(W^{*}_{k})]-\mathrm{E}[g_{s}(\tau_{1}+W^{*}_{k})] \,e^{-\alpha_{k}\tau_{1}} \right\rbrace \\
\\
&&-\displaystyle \sum_{j=1}^{4}\,p^{(1)}_{ij}(\tau_{1})\, \displaystyle\sum_{k:\beta_{k}>0}C_{jk}^{-1}C_{k4}\left\lbrace\mathrm{E}[g_{s}(\tau_{1}+Y^{*}_{k})]\right.\\
\\
&&\left. -\mathrm{E}[g_{s}(\tau_{0}+Y^{*}_{k})]\, e^{-\beta_{k}(\tau_{0}-\tau_{1})} \right\rbrace\\
\\
&&-  \displaystyle\sum_{j=1}^{4} p^{(1)}_{ij}(\tau_{1})\displaystyle\sum_{l=1}^{4}p^{(2)}_{jl}(\tau_{0}-\tau_{1})\,\displaystyle\sum_{k:\gamma_{k}>0} D_{lk}^{-1}D_{k4}\,\mathrm{E}[g_{s}(\tau_{0}+Z^{*}_{k})] \quad.
\end{array}
\end{equation} 
\normalsize
 To give an explicit statement of the expectations in this probability mass function, we use the results of equations (16) and (17) in \citet{Herbots2012}: for a random variable $U$ following an exponential distribution with rate $\lambda$,
\small
\begin{equation}
\label{eq:int_1}
\begin{array}{ll}
\mathrm{E}[g_{s}(U)]&=\left(\frac{\theta}{\lambda+\theta}\right)^{s} \left(\frac{\lambda}{\lambda+\theta}\right)\quad
\end{array}
\end{equation} 
\normalsize

\noindent
and 
\small
\begin{equation}
\label{eq:int_2}
\begin{array}{ll}
\mathrm{E}[g_{s}(\tau+U)]&=\left(\frac{\theta}{\lambda+\theta}\right)^{s} \left(\frac{\lambda}{\lambda+\theta}\right)\,e^{-\theta \tau}\sum_{l=0}^{s}\frac{\left(\lambda+\theta\right)^{l}\tau^{l}}{l!} \quad.
\end{array}
\end{equation}
\normalsize

\subsection{The likelihood of a multilocus data set}
Recall that, for the purposes of this paper, an observation consists of the number of nucleotide differences between two DNA sequences at a given locus.  To fit the GIM model, we need a large set of observations from each of the three possible initial states: both sequences sampled from species 1 (state 1); both sequences sampled from species 2 (state 2); and one sequence from each species (state 3). To compute the likelihood of such a set, we make use of the assumption of free recombination between loci.

Let $\boldsymbol{\rho}$ be the vector of parameters of the coalescent under the GIM model, i.e. 
\small
\begin{equation*}
\boldsymbol{\rho}=[a \quad b \quad c_{1} \quad c_{2} \quad\tau_{1} \quad \tau_{0} \quad M_{1} \quad M_{2} \quad M'_{1} \quad M'_{2}] \quad.
\end{equation*}
\normalsize
Furthermore, let $\theta$ now denote the average mutation rate over all loci in the data set, and let the mutation rate at a given locus $l$ be represented by $\theta_{l}$. The parameter $\theta_{l}$ can be written as $\theta_{l}=r_{l}\theta$, where $r_{l}=\frac{\theta_{l}}{\theta}$ is the \textit{relative} mutation rate of locus $l$. If the $r_{l}$ are known, the likelihood of a set of observations from  independent loci can be written as
\small
\begin{equation*}
\label{eq:estimated likelihood}
\displaystyle 
L\left(\boldsymbol{\rho},\theta;\mathbf{x},\mathbf{r}\right)=\prod_{l} L(\boldsymbol{\rho},\theta;x_{l},r_{l}) \quad,
\end{equation*}
\normalsize
where  $L(\boldsymbol{\rho},\theta;x_{l},r_{l})$, the likelihood of the observation from locus $l$, has the same form as equation (\ref{eq:likelihood_single}), but with  $\theta$ replaced by $r_{l}\theta$ in equations (\ref{eq:int_1}) and (\ref{eq:int_2}).

For real data sets, the relative mutation rates must be estimated and substituted into the likelihood before any inference can be carried out. Estimates of $r_{l}$ can be computed by means of the following estimator suggested by \cite{Yang2002}, in which $L$ is the total number of loci, and $\bar{d}_{l}$ is the average, at locus $l$, of the ingroup-outgroup pairwise distance estimates (i.e the average is over all the distance estimates that can be computed at locus $l$ using  pairs of sequences that are composed of one ingroup sequence and one outgroup sequence):
\small
\begin{equation*}
\hat{r}_{l}=\frac{L\hspace{0.1cm}\bar{d}_{l}}{\sum^{L}_{m=1}\bar{d}_{m}}\qquad.
\end{equation*}
\normalsize

\section{Discussion}

The main aim of this paper is to enable the comparison of three different scenarios for the divergence of closely related pairs of species (divergence without gene flow, with ancestral gene flow followed by isolation, and with continuous gene flow until the present), in a setting that allows for population sizes and migration rates to change during the divergence process. We achieve this aim by developing a maximum-likelihood method to fit the models illustrated in Figure \ref{fig:GIM}  to DNA sequence data sets. A formal comparison of the different versions of the GIM model, by means of likelihood ratio tests or AIC scores, can easily be carried out. In \citet{Herbots2015} and \citet{Costa2016}, we show how to implement this sort of model selection procedure for the isolation-with-initial-migration model.

\begin{figure}[h] 
\graphicspath{ {./} }
\centering         
\includegraphics[width=7.5cm, angle=0]{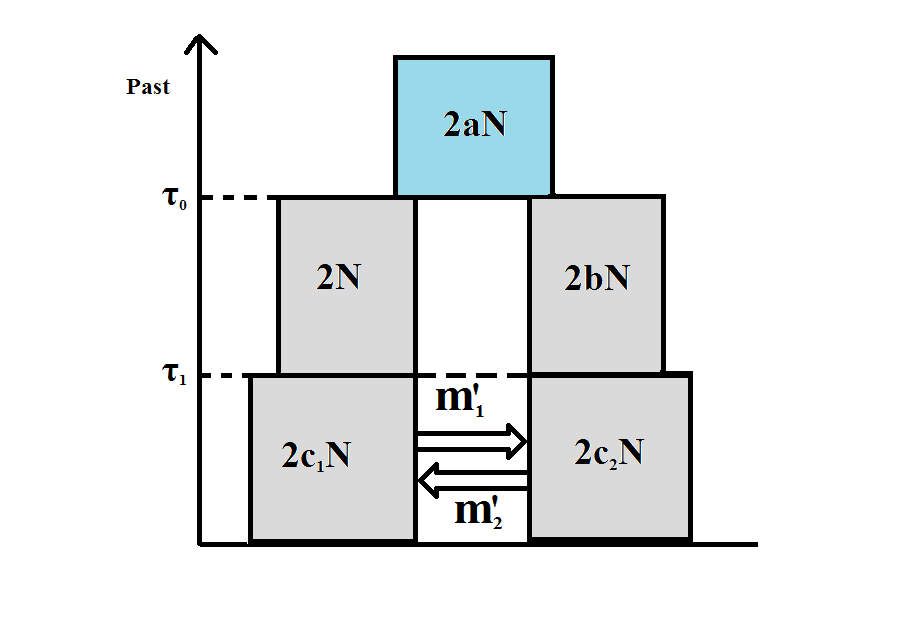} 
\vspace*{-0.25cm}     
\caption{A model of divergence in which current gene flow is preceded by a period of isolation (a GIM model with $m_{1}=m_{2}=0$).}      
\label{fig:invIIM} 
\end{figure}

 The likelihood given in equation (\ref{eq:likelihood_single}) allows the estimation of the GIM model (see Figure \ref{fig:GIM}, central diagram) and any model nested in it, including models with a single divergence stage, such as the complete isolation and the IM models represented in Figure \ref{fig:IIMnested}. 
A special case of the GIM model which may be of particular interest represents a scenario of introgression as illustrated in Figure \ref{fig:invIIM}, where gene flow occurs between two diverging species after a period of isolation. Such a scenario may have been caused, for example, by climatic changes leading to habitat fragmentation and subsequent reconnection of populations.

The extension of the present method to the Jukes-Cantor model of mutation should be relatively straightforward. Under this model of mutation, the probability mass function of the number of pairwise differences given $T$, the coalescence time, can be written as the sum of moment generating functions of pairwise coalescence times \citep[see][equation (3)]{Lohse2011}. Hence integrating out $T$ analytically is still possible. This is left for future work.

\section*{Acknowledgements}

This research was supported by the Engineering and Physical Sciences Research Council (grant number EP/K502959/1).

\clearpage
\bibliographystyle{Chicago}
\bibliography{GIMpaper}
\end{document}